\documentclass[aps,prb,twocolumn,preprintnumbers,groupedaddress,showpacs]{revtex4}

\usepackage{graphicx}
\usepackage{dcolumn}
\usepackage{bm}

\begin{document}
\newcommand {\etal}{{\it et al.}~}
\bibliographystyle{apsrev}

\title{Quantum dynamics of crystals of molecular nanomagnets inside a
resonant cavity}

\author{J. Tejada\email{jtejada@ubxlab.com}}
\affiliation{Department de F{\'\i}sica Fonamental, Universitat de
Barcelona, Diagonal 647, 08024 Barcelona, Spain}

\author{E. M. Chudnovsky}
\affiliation{Department of Physics and Astronomy, Lehman College,
City University of New York, 250 Bedford Park Boulevard West
Bronx, New York 10468-1589, USA}

\author{R. Amigo}
\author{J. M. Hernandez}
\affiliation{Department de F{\'\i}sica Fonamental, Universitat de
Barcelona, Diagonal 647, 08024 Barcelona, Spain}

\begin{abstract}
  It is shown that crystals of molecular nanomagnets exhibit
enhanced magnetic relaxation when placed inside a resonant cavity.
Strong dependence of the magnetization curve on the geometry of
the cavity has been observed, providing evidence of the coherent
microwave radiation by the crystals. A similar dependence has been
found for a crystal placed between Fabry-Perot superconducting
mirrors. These observations open the possibility of building a
nanomagnetic microwave laser pumped by the magnetic field.

\end{abstract}

 \maketitle

\begin{figure}[t]
\centering
\includegraphics[width=3in]{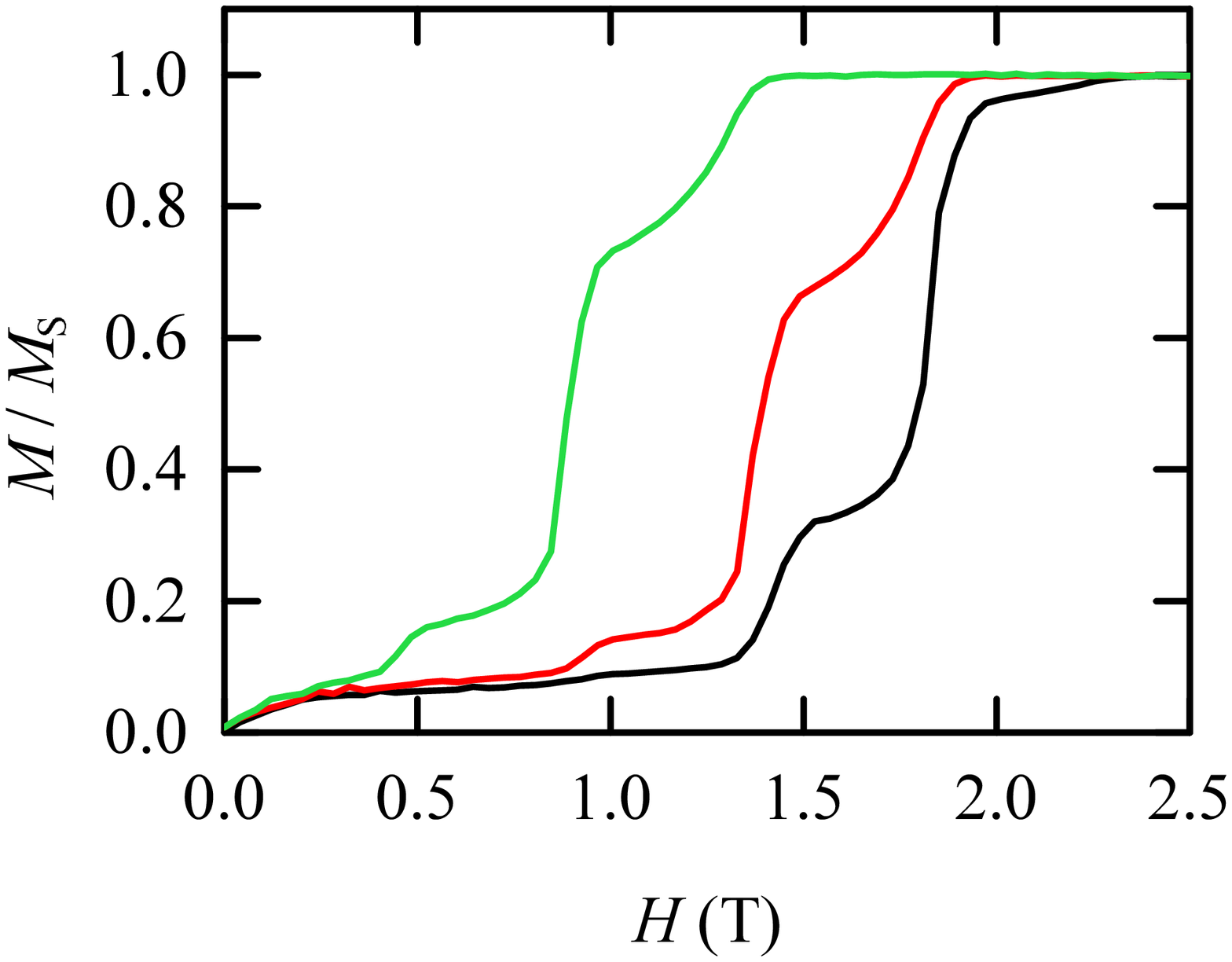}
\caption{ \label{fig_qmc} Initial magnetization curves of
Mn$_{12}$-acetate at $T=1.8\,$K (black), $T=2.0\,$K (red), and
$T=2.4\,$K (green).}
\end{figure}

\begin{figure}[t]
\centering
\includegraphics*[width=2in]{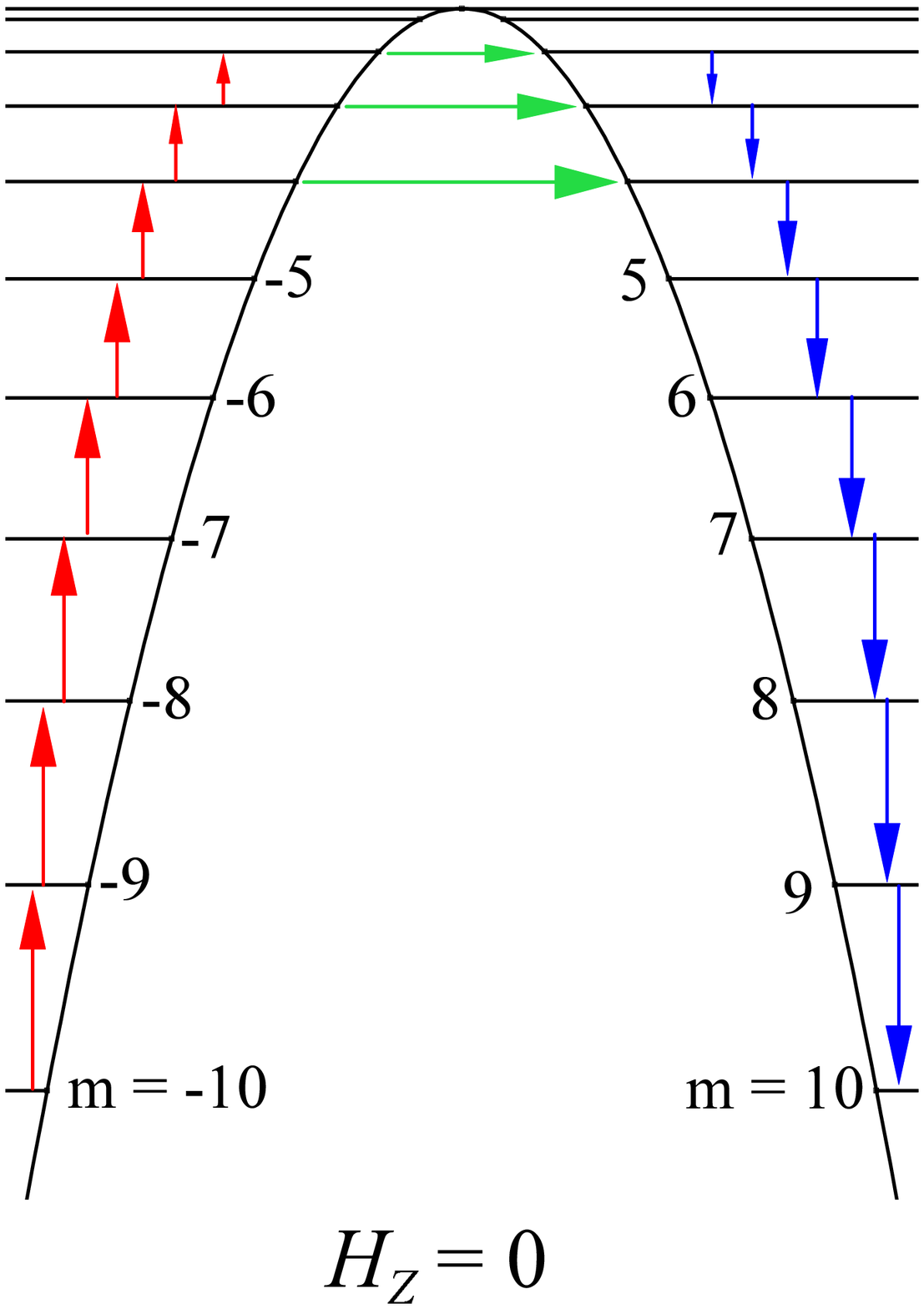}
\caption{ \label{fig_levels} Thermally assisted spin tunneling in
spin-10 molecular magnets.}
\end{figure}

Many fascinating magnetic effects occur at the boundary between
classical and quantum physics of the angular momentum. They have
been intensively investigated, both theoretically and
experimentally, over the last two decades \cite{book}. The
discovery of high-spin molecular nanomagnets has given a strong
boost to that field. The most carefully studied nanomagnets are
Mn$_{12}$ and Fe$_{8}$ molecular clusters. Both have spin 10,
which is twenty times the spin of electron. Such a large angular
momentum is almost classical. Still crystals of Mn$_{12}$ and
Fe$_{8}$ exhibit spectacular magnetic effects related to the
quantization and quantum tunneling of the angular momentum. The
current great interest of physicists to these systems was ignited
by the discovery of the quantum stepwise magnetization curve in
Mn$_{12}$-acetate \cite{four}, see Fig.\ \ref{fig_qmc}. It has the
following simple explanation. The Mn$_{12}$ cluster has a
tetragonal symmetry \cite{Roberta}. In the magnetic field $H_{z}$
applied along the tetragonal axis, it is described by the
Hamiltonian
\begin{equation}\label{Hamiltonian}
{\cal{H}}=-DS_z^2-AS_z^4-g{\mu}_{B}S_{z}H_{z}+{\cal{H}}',
\end{equation}
where
$S=10$,$\;D\,{\approx}\,0.55\,$K,$\;A\,{\approx}\,1.2{\times}10^{-3}$K,
$\;g\,{\approx}\,1.94$, and ${\cal{H}}'$ contains small terms that
do not commute with $S_{z}$. If one neglects ${\cal{H}}'$, then,
in a zero field, the ground state of the cluster is double
degenerate. The two energy minima correspond to the spin looking
up or down along the $Z$-axis. They are separated by the magnetic
anisotropy barrier $U\,{\approx}\,65\,$K. At a non-zero field, the
trivial algebra of Eq.\ (\ref{Hamiltonian}) yields that the $m$
and $m'$ eigenvalues of $S_{z}$ come to resonance at
\begin{equation}\label{resonances}
H_{z}=H_{mm'}=-(m+m')\frac{D}{g{\mu}_{B}}\left[1+\frac{A}{D}(m^{2}+{m'}^{2})\right]\;,
\end{equation}
see Fig.\ \ref{fig_levels}. At $H_{z}\,{\neq}\,H_{mm'}$
transitions between negative and positive $m$ occur due to the
thermal activation over the anisotropy barrier in Fig.\
\ref{fig_levels}. However, at $H_{z}=H_{mm'}$ the transitions are
combinations of thermal activation and quantum tunneling
\cite{Leo,Rolf,CG,Garanin91}. Thus, for $H_{z}=H_{mm'}$ the
anisotropy barrier is effectively reduced by tunneling and the
magnetic relaxation towards the direction of the field is faster
than at $H_{z}\,{\neq}\,H_{mm'}$. This picture, first established
by magnetic measurements \cite{four}, has been confirmed by the
careful EPR study of spin levels in Mn$_{12}$-acetate
\cite{Barra-ESR,Hill,Javier,Sushkov-ESR,Parks-ESR,Park-ESR,Dressel}.

The EPR experiments have demonstrated noticeable resonant
absorption of electromagnetic radiation by molecular magnets. Most
recently, it has been suggested that a crystal of magnetic
molecules can also be a powerful source of coherent microwave
radiation \cite{CG-SR}. The radiation can be produced during the
relaxation of the crystal towards the minimum of the magnetic
energy. In this paper we provide first experimental support to
this suggestion by demonstrating strong interaction between the
Mn$_{12}$-crystal and a microwave cavity. Our intention was to
test whether placing the crystal inside a resonant cavity would
result in any change in the magnetization curve. Before showing
the experimental data, let us discuss why the detection of such an
effect would constitute the evidence of the coherent radiation
inside the cavity.

Consider a crystal of Mn$_{12}$-acetate magnetized in the negative
$Z$-direction. At $|H_{z}|<<10\,$T the distance $E_{1}$ between
$m=-10$ and $m=-9$ levels in Fig.\ \ref{fig_levels} is about
$14\,$K. Thus, at $T<<E_{1}$, most of the molecules are occupying
the $m=-10$ level. The occupation numbers of the excited states
scale as $\exp(-E/T)$ where $E$ is the energy distance from
$m=-10$. In the absence of a very large transverse field, quantum
tunneling from $m=-10$ in Mn$_{12}$ has a negligibly low
probability. The thermodynamic equilibrium is achieved either
through thermal activation or through thermally assisted quantum
tunneling from excited states. The activation of molecules to the
excited states is believed to be due to the absorption of phonons.
At $T<<E_{1}$ the rate of phonon-induced transitions from $m=-10$
to $m=-9$ for the Hamiltonian of Eq.\ (\ref{Hamiltonian}) is given
by \cite{phonons}
\begin{equation}\label{phonon}
{\Gamma}_{phonon}=\frac{{\hbar}S{\omega}_{1}^{5}}{12{\pi}{\rho}c_{s}^{5}}
\exp\left(-\frac{{\hbar}{\omega}_{1}}{T}\right)\;,
\end{equation}
where ${\omega}_{1}=E_{1}/{\hbar}$ is the frequency of the phonon
($f_{1}={\omega}_{1}/2{\pi}\,{\approx}\,300\,$GHz),
${\rho}\,{\sim}\,1.8\,$g/cm$^{3}$ is the mass density of the
crystal, and $c_{s}\,{\sim}\,10^{5}\,$cm/s is the speed of the
transverse sound. At $2\,$K Eq.\ (\ref{phonon}) yields
${\Gamma}_{phonon}\,{\sim}\,5{\times}10^{5}$s$^{-1}$.

In order to have any effect due to the electromagnetic radiation,
the absorption of photons must have the rate comparable to the
rate of the absorption of phonons. This can be achieved in the EPR
experiment in which the sample is placed in the ac magnetic field
$H_{ac}$ oscillating at frequency $f_{1}$. The rate of the
absorption of photons in the EPR setup is given by \cite{Abragam}
\begin{equation}\label{Abragam}
{\Gamma}_{photon}=k\frac{Sg^{2}{\mu}_{B}^{2}}{{\hbar}^{2}}H_{ac}^{2}F(\omega)\;,
\end{equation}
where $k$ is a numerical factor of order one that depends on the
polarization of photons and $F(\omega)$ is the shape function of
the resonance. In the case of the Lorentzian line of the width
${\Delta}{\omega}$, the shape function is given by
\begin{equation}\label{Lorentzian}
F(\omega)=\frac{1}{\pi}\frac{{\Delta}{\omega}}{({\Delta}{\omega})^{2}+({\omega}-{\omega}_{1})^{2}}\;.
\end{equation}
It reduces to the Delta-function ${\delta}({\omega}-{\omega}_{1})$
for ${\Delta}{\omega}\,{\rightarrow}\,0$. At a finite width, the
maximal EPR rate is achieved at ${\omega}={\omega}_{1}$ and is
given by
${\Gamma}_{photon}\,{\sim}\,g{\mu}_{B}SH_{ac}^{2}/{\hbar}{\Delta}H$,
where ${\Delta}H$ is the field width of the line,
${\hbar}{\Delta}{\omega}=g{\mu}_{B}{\Delta}H$. In Mn$_{12}$,
$\;{\Delta}H$ is of order of $400\,$Oe. Consequently, at
$H_{ac}\,{\sim}\,1\,$Oe the rate of the absorption of EPR photons
becomes comparable to the phonon rate of Eq.\ (\ref{phonon}) and
the microwave power delivered to the Mn$_{12}$ crystal can
significantly alter the magnetic relaxation.

In our experiments no external ac magnetic field has been used.
The crystal of Mn$_{12}$-acetate was simply placed inside a
resonant cavity and the magnetization curve has been measured. Let
us assume for the moment that only thermal photons are available
for the transitions between, e.g., $m=-10$ and $m=-9$ levels. The
wavelength of these photons is comparable to the dimensions of the
cavity. Their magnetic field can be then estimated from
$H_{ac}^{2}/8{\pi}\,{\sim}\,({\hbar}{\omega}_{1}/V)\exp(-{\hbar}{\omega}_{1}/T)$,
where $V\,{\sim}\,5{\times}10^{-2}$cm$^{3}$ is the volume of the
cavity. This gives $H_{ac}\,{\sim}\,3{\times}10^{-8}$Oe, as
compared to $H_{ac}\,{\sim}\,1\,$Oe needed to beat the phonon rate
in the EPR experiment at ${\Delta}H\,{\sim}\,400\,$Oe. Thus at
$2\,$K thermal photons inside the cavity excite Mn$_{12}$
molecules at a rate that is fifteen orders of magnitude lower than
the phonon rate. This is mainly due to the fact that at any
temperature each cubic centimeter of a solid contains
$(c/c_{s})^{3}\,{\sim}\,10^{15}$ times more thermal phonons than
thermal photons. At $T=2\,$K our cavity would have $10^{-3}$
average number of thermal photons of energy $E_{1}$. Consequently,
the cavity should play absolutely no role in the magnetic
relaxation unless it occasionally acquires a very large number of
non-thermal photons. If all these photons have the same phase,
their effect will be equivalent to the effect of the ac magnetic
field in the EPR experiment and may become comparable to the
effect of the phonons.

Single crystals of Mn$_{12}$-acetate have been grown with the
average length and diameter of about $2\,$mm and $0.2\,$mm
respectively. The elongation of the crystals was along the c-axis.
The conventional composition and the structure of the crystals
\cite{Lis} have been established by chemical, infrared, and X-ray
diffraction techniques. In addition, the dc and ac magnetometry of
the crystals have been performed. The same values of the blocking
temperature and resonance fields, as previously reported
\cite{four,Hernandez,Barbara}, have been found.

In constructing resonant cavities we followed the procedures
described in Refs.\cite{Steve,Martin}. Five cylindrical cavities
of different diameter and adjustable length were constructed using
$99.99$\% purity copper. Two diameters were used: $1.6\,$mm and
$3.2\,$mm. The length of the cavity was controlled with the help
of the same-purity copper rod connected to the upper surface of
the cavity. The micrometric stepping motor control system was used
that had the spatial resolution of $1\,$${\mu}$m. The inner
lateral and planar surfaces of the cavity were polished to a
roughness of less than $10\,$${\mu}$m to achieve the quality
factor between $Q=10^{3}$ and $Q=10^{4}$. The crystals were fixed
in a vertical position at the bottom of the cavities, with the
c-axis axis of the crystal parallel to the cylindrical axis of the
cavity. The grease of high thermal conductivity was used to attach
the crystal to the bottom of the cavity. The cavity containing the
crystal was immersed in a helium gas inside the MPMS SQUID
magnetometer. The upper moving surface made the cavity not
hermetic, thus allowing the exchange of helium with the outside
reservoir. The temperature inside and outside the cavity was
monitored by carbon thermometers.
\begin{figure}[t]
\centering
\includegraphics[width=3in]{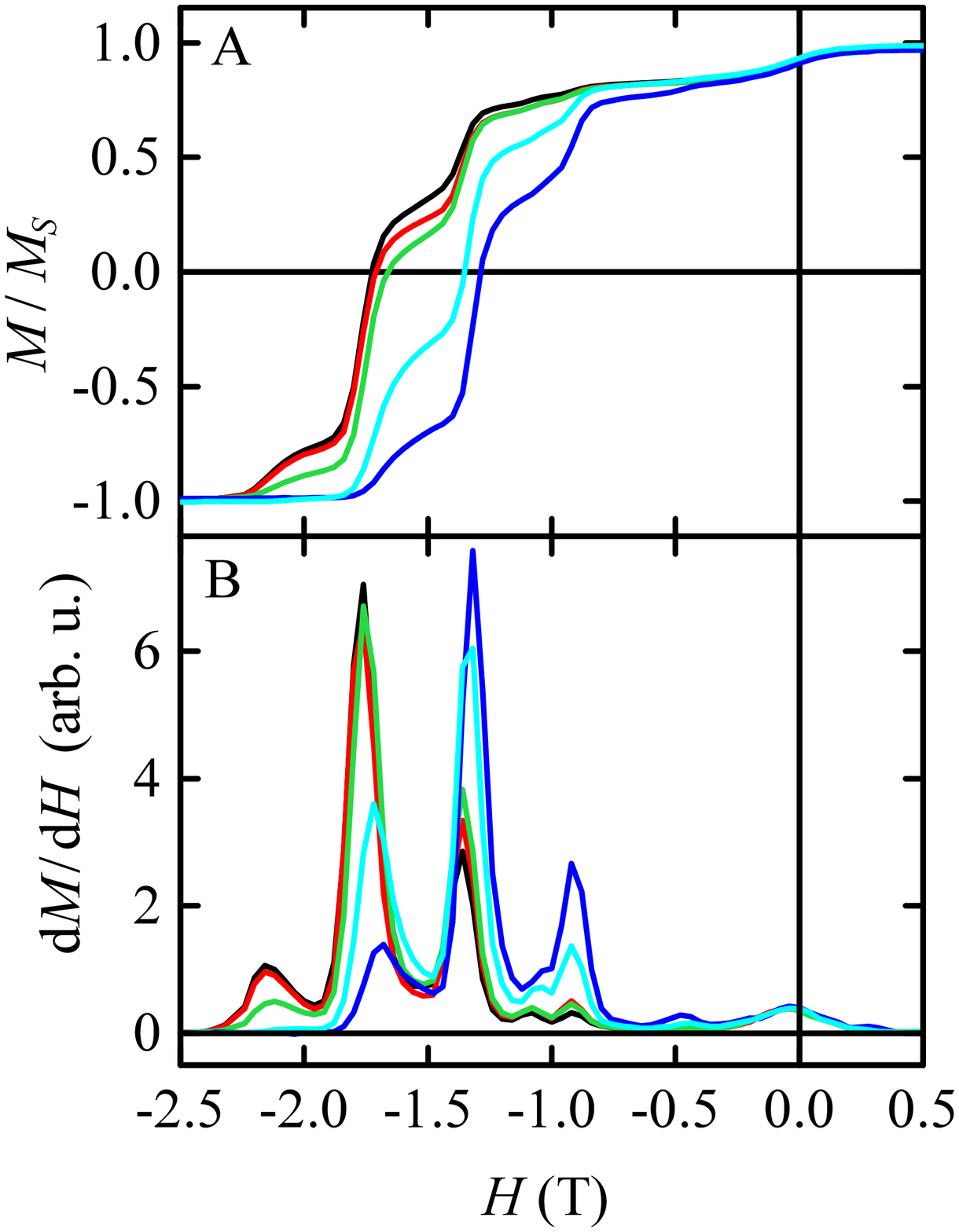}%
\caption{ \label{fig_cavity} (A) Demagnetization curves, $M(H)$,
of Mn$_{12}$-acetate crystal at $T=2.0\,$K inside the resonant
cavity of diameter $1.6\,$mm for five different lengths of the
cavity: $L=21.0\,$mm (black), $L=20.9\,$mm (red), $L=20.1\,$mm
(green), $L=19.8\,$mm (dark blue), and $L=19.5\,$mm (light blue);
(B) $dM/dH$ for the curves shown in (A).}
\end{figure}

\begin{figure}[t]
\centering \includegraphics[width=3in]{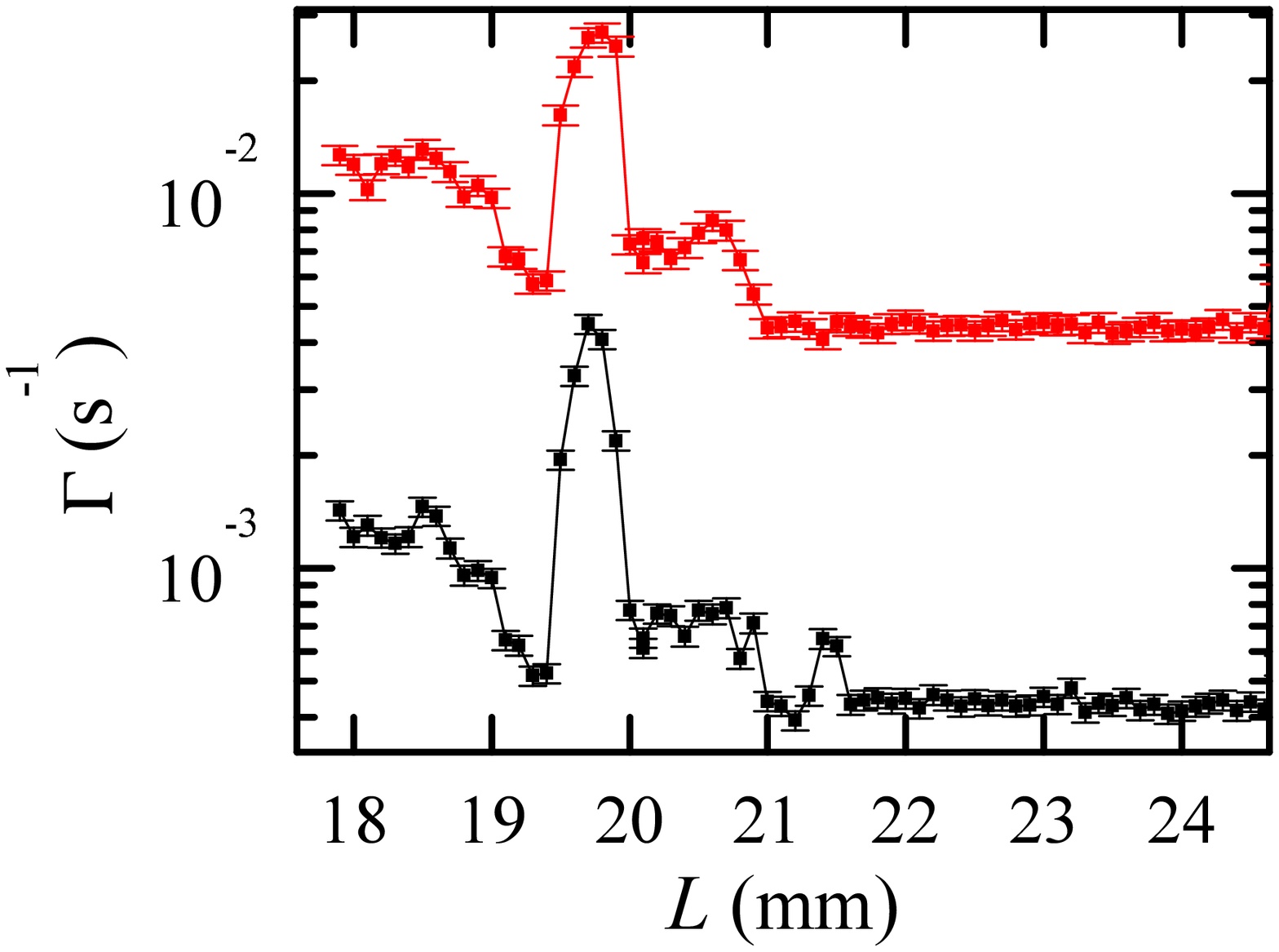}%
\caption{ \label{fig_length} The rate of magnetic relaxation in a
Mn$_{12}$-acetate crystal inside the cavity of diameter $1.6\,$mm
at $T=2.0\,$K as a function of the cavity length at the second
(black points) and third (red points) resonance fields.}
\end{figure}

Before placing crystals inside the cavities the magnetic signal
from the crystals and from the cavities have been measured
independently. The signal from the crystal was always two orders
of magnitude greater than the paramagnetic signal from the cavity.
We then proceeded to the measurements of crystals inside the
cavities. The system was first saturated by a $5\,$T field applied
along the c-axis of the crystal. Then the field was swept in the
opposite direction at the rate $40\,$Oe/s. Very similar behavior
has been found in all of the cavities. Typical demagnetization
curves from Mn$_{12}$-acetate crystals inside a resonant cavity
are shown in Fig.\ \ref{fig_cavity}. The dependence of the
relaxation rate, ${\Gamma}=|M(H)-M_{eq}(H)|^{-1}(dM/dH)$ (with
$M_{eq}$ being equilibrium magnetization), on the length of the
cavity at two resonance fields is shown in Fig.\ \ref{fig_length}.

\begin{figure}[t]
\centering
\includegraphics[width=3in]{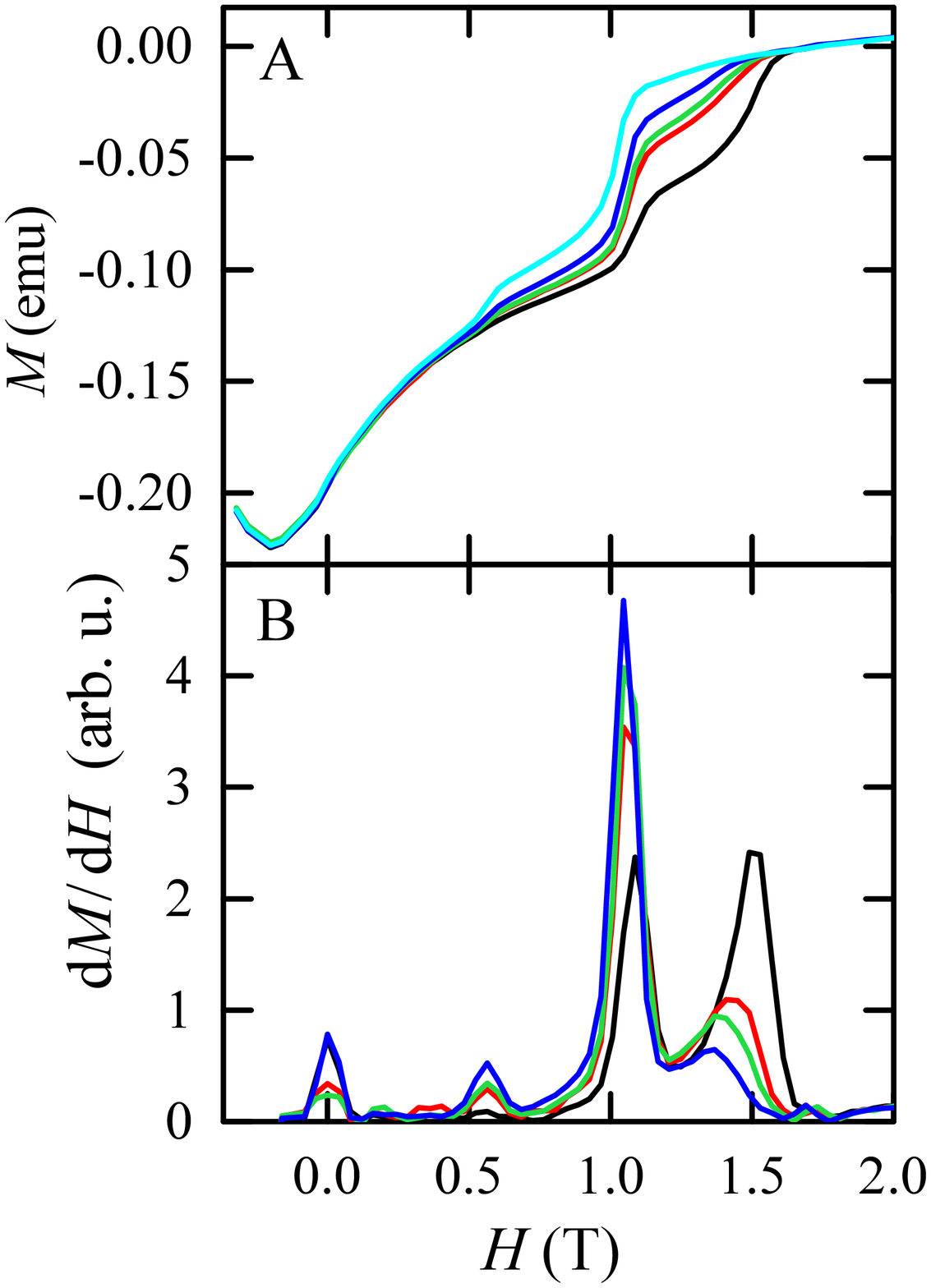}%
\caption{ \label{fig_mirrors} (A) Demagnetization curves, $M(H)$,
of the Fabry-Perot setup at $T=2.0\,$K for five distances between
superconducting mirrors: $d=4.3\,$mm (black), $d=4.7\,$mm (red),
$d=5.1\,$mm (green), $d=5.3\,$mm (dark blue), and $d=6.0\,$mm
(light blue); (B) $dM/dH$ calculated from the demagnetization
curves in (A) after subtracting the contribution of the
superconductors.}
\end{figure}

\begin{figure}[t]
\centering
\includegraphics[width=3in]{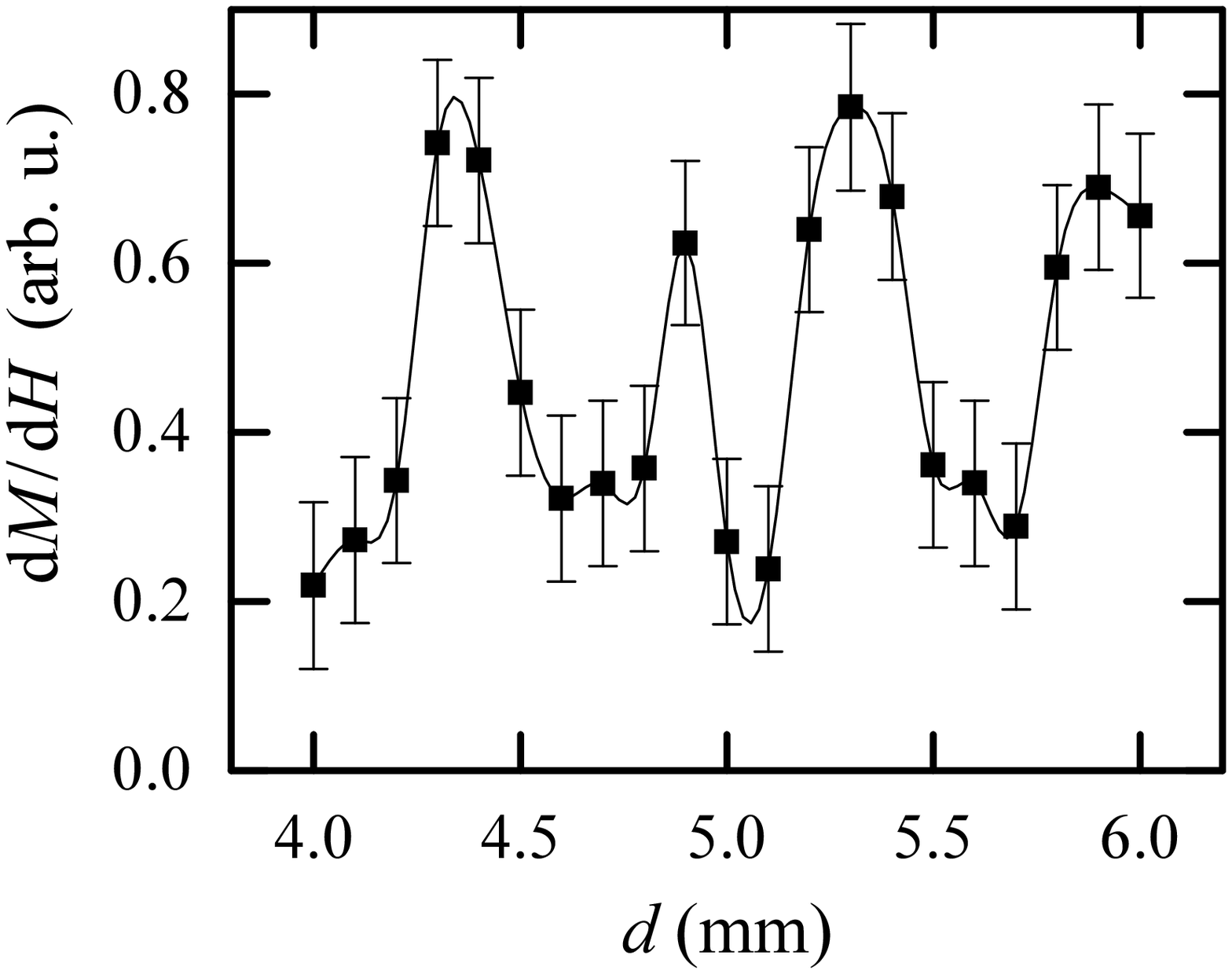}%
\caption{ \label{fig_mirrors-L} $dM/dH$ at $H_{z}=0$ and
$T=2.0\,$K for a Mn$_{12}$-acetate crystal between superconducting
mirrors as a function of the distance between the mirrors.}
\end{figure}

In a separate set of experiments we placed the Mn$_{12}$-acetate
crystal between two Fabry-Perot superconducting mirrors. The
mirrors were prepared using method described in Ref.
\cite{mirrors}. The $200\,$nm YBaCuO layers were deposited by
pulsed laser deposition on a $1\,$${\mu}$m SrTiO$_3$ substrate.
Their superconducting properties below $90\,$K were verified by
magnetic measurements. The magnitude of the diamagnetic signal
from the mirrors was found to be comparable to the magnitude of
the signal from the crystal, but independent from the distance
between the mirrors. The demagnetization curves from the
Fabry-Perot setup are shown in Fig.\ \ref{fig_mirrors}. The
diamagnetic signal from the superconductors was subtracted from
the total signal to obtain Fig.\ \ref{fig_mirrors}B. The
dependence of $dM/dH$ on the distance between the mirrors at
$H_{z}=0$ is shown in Fig.\ \ref{fig_mirrors-L}.

The experimental data clearly demonstrate the dependence of the
magnetic relaxation in Mn$_{12}$-acetate crystals on the geometry
of the cavity. We have verified that this phenomenon appears only
in cavities of high quality factor. Could it be due to effects
unrelated to the electromagnetic properties of the cavity? The
first thing that comes to mind is that in the confined geometry
the cooling of the crystals by the flow of helium could be not
perfect. The rate of thermally assisted quantum tunneling in
Mn$_{12}$ depends on temperature exponentially \cite{Hernandez}.
Thus, fluctuations in the hermetic properties of the cavity could,
in principle, result in a significant change of the relaxation
rate. For that reason we carefully monitored temperature inside
the cavity. The temperature variation was found not to exceed
$0.3\,$\%. Such small fluctuations of temperature cannot account
for any measurable change in the relaxation rate. All data on the
dependence of the rate on the length of the cavity and the
magnetic field were reproducible within a $6\,$\% experimental
error. Strong dependence of the relaxation rate on the geometry of
the confinement was also observed in a Mn$_{12}$-acetate crystal
placed between superconducting mirrors, that is, when the helium
gas circulated freely around the crystal. We, therefore, concluded
that the observed phenomenon had nothing to do with thermal
effects and was, in fact, due to the microwave properties of the
cavity.

As has been discussed above, in order to affect the magnetic
relaxation of a Mn$_{12}$ crystal, the number of coherent photons
in the cavity must be very large compared to the number of thermal
photons. There is only one source of these photons, it is the
crystal itself. To change the orientation of the magnetic moment,
the molecule must first go up the staircase of spin levels and
then go down (see Fig.\ \ref{fig_levels}). Thus, the magnetic
relaxation in the crystal creates massive inversed population of
the energy levels. These excited states decay via emission of
phonons or photons. Without a cavity, the corresponding
magnetic-dipole photon transitions would have a negligible
probability compared to the probability of phonon transitions.
Consequently, the relaxation towards thermal equilibrium would
occur via the emission of phonons. Inside the cavity, however, a
maser effect can take place if some of the frequencies of the
emitted photons coincide with resonances of the cavity. The photon
emitted by one molecule remains in the cavity and stimulates the
emission of photons by other molecules. The shortest wavelength,
${\lambda}$, of a photon corresponds to the transition from $m=9$
to $m=10$. For Mn$_{12}$ at $H=0$ it is about $1\,$mm. The
wavelengths of photons emitted in other transitions are longer.
Consequently, for a crystal of size $2\,$mm, the phase of the
emitted photons is the same for a macroscopically large number of
molecules, $N=N_{SR}\,{\sim}\,({\lambda}/2)^{3}$. In that case the
emission of photons by different molecules becomes correlated and
the superradiance may occur \cite{CG-SR,SR}: The rate of the
emission of a photon increases by a factor $N_{SR}$. This effect
is much stronger for photons than for phonons because the
wavelength of photons is $(c/c_{s})$ times the wavelength of
phonons of the same energy. Thus, the $(c_{s}/c)^{3}$ smallness of
the phase space of photons in comparison with the phase space of
phonons is compensated by the $(c/c_{s})^{3}$ times greater
$N_{SR}$ for photons as compared to phonons. Crystals used in our
experiments contained about $1.6{\times}10^{16}$ Mn$_{12}$
molecules, so that each magnetization step in Fig.\
\ref{fig_cavity} and Fig.\ \ref{fig_mirrors} involved more than
$10^{15}$ molecules. For $Q\,{\sim}\,3{\times}10^{3}$ the
$300\,$-Ghz photons stay inside the cavity during
$t\,{\sim}\,10\,$ns. The reabsorption of the radiation by the
crystal requires ${\Gamma}_{photon}t>1$. According to Eq.\
(\ref{Abragam}), our picture is self-consistent if during the
$10\,$s field sweep across the tunneling resonance of width
${\Delta}H\,{\sim}\,400\,$Oe the cavity was affecting the magnetic
relaxation through $10^{3}({\Delta}H/{\Delta}h)$ microbursts of
superradiance of duration $10\,$ns or less, each burst involving
$10^{12}({\Delta}h/{\Delta}H)$ molecules whose resonances were
distributed within ${\Delta}h<1\,$Oe.

The effect of $10^{12}({\Delta}h/{\Delta}H)$ coherent photons
inside the cavity would be similar to the effect of the $1\,$Oe ac
magnetic field in the EPR experiment. That is, the absorption of
the photons increases the population of the levels with $m>-10$ in
Fig.\ \ref{fig_levels} as compared to their thermal population.
Contrary to the EPR experiment, however, in our case the photons
absorbed in the left well in Fig.\ \ref{fig_levels} must be the
same photons that are emitted in the right well. One can think
about this process as the recycling of the emitted photons by the
crystal after the number of photons in the cavity reaches the
critical value. To affect the magnetic relaxation, one of the
frequencies, $\omega$, of the emitted photons should satisfy two
conditions. The first condition is that ${\hbar}{\omega}$
coincides with one of the distances between the spin levels in the
metastable well (the left staircase in Fig.\ \ref{fig_levels}). It
is fulfilled for certain values of the field $H_{z}=H_{r}$ which
are uniquely determined by the spin Hamiltonian. If $A$ in Eq.\
(\ref{Hamiltonian}) was zero, the fields $H_r$ would coincide with
$H_{mm'}(A=0)=H_{n}=nD/g{\mu}_{B}$. Because of $A\,{\neq}\,0$,
however, $H_{r}$ are different from $H_{mm'}$ of Eq.\
(\ref{resonances}), except for $H_{z}=0$. Nevertheless, due to the
fact that $A<<D$, the fields $\;H_{r}$ must group around $H_{n}$.
The second condition is that the resonant frequency corresponding
to $H_{z}=H_{r}$ also coincides with one of the resonances of the
cavity. It is achieved by manipulating the length of the
cylindrical cavity or the distance between superconducting mirrors
in the Fabry-Perot setup.

Strong support to the above picture comes from the fact that the
main period of oscillations of the relaxation rate on the distance
between superconducting mirrors, (see Fig.\ \ref{fig_mirrors-L})
is about $0.5\,$mm, which is one-half of the wavelength of the
$300\,$Ghz photons responsible for the transitions between $m=-10$
and $m=-9$ levels. This resonance is the most important one
because almost all molecules initially occupy the $m=-10$ state.
Exciting these molecules to the $m=-9$ level alone changes the
effective energy barrier by $E_{1}\,{\approx}\,14\,$K, thus
increasing the relaxation rate by a factor
$\exp(E_{1}/T)\,{\sim}\,10^{3}$. For cavities, the
length-dependence of the relaxation rate is more complicated.
Photons emitted in the magnetic dipole transitions should be in
resonance with the TM modes of the cavity. The latter, for a
cavity of radius $R$ and length $L$, satisfy
\begin{equation}\label{cavity}
{\omega}^{2}_{mnp}=\frac{c^{2}{\kappa}^{2}_{mn}}{R^{2}}+\frac{{\pi}^{2}c^{2}p^{2}}{L^{2}}\;,
\end{equation}
where $p=1,2,3,...$ and $x={\kappa}_{mn}$ is the n-th zero of the
Bessel function $J_m(x)$. Matching the spectrum of the spin levels
of Mn$_{12}$-acetate with the spectrum of the cavity should be the
way to explain the dependence of the magnetic relaxation on the
length of the cavity. We have not succeeded in that task so far.
Both spectra are rather dense. Their comparison is complicated by
the distribution of energy levels due to dipolar and hyperfine
fields, crystal imperfections, etc.

The main message that we want to disseminate is that the magnetic
relaxation of molecular nanomagnets inside a resonant cavity
differs from the magnetic relaxation outside the cavity. It seems
impossible to understand this effect without invoking the
superradiance. Our findings open the possibility of building a
microwave laser pumped by the magnetic field. Measurements of
microwave radiation from cavities containing crystals of molecular
nanomagnets should be the next step in this direction.

We thank F. S\'{a}nchez and M. Varela for help in preparation of
Fabry-Perot superconducting mirrors. The financial support from
the EU is gratefully acknowledged. The work of E.M.C. has been
supported by the U.S. NSF Grant No. 9978882.

\end{document}